\documentclass[aps,superscriptaddress,twocolumn,showpacs,amsmath,amssymb]{revtex4}
\usepackage{graphicx}
\usepackage{dcolumn}
\usepackage{bm}
\usepackage{amsmath,amssymb,amsfonts}

\begin{document}

\title{Effects of mode-mode and isospin-isospin correlations
on domain formation of disoriented chiral condensates}

\author{N. Ikezi}
\affiliation{
Department of Physics, Osaka University, Toyonaka 560-0043, Japan}
\affiliation{
Research Center for Nuclear Physics (RCNP),
Osaka University, Ibaraki 567-0047, Japan.}

\author{M. Asakawa}
\affiliation{
Department of Physics, Osaka University, Toyonaka 560-0043, Japan}

\author{Y. Tsue}
\affiliation{
Physics Division, Faculty of Science, Kochi University, Kochi
780-8520, Japan}

\date{\today}

\begin{abstract}
The effects of mode-mode and isospin-isospin
correlations on nonequilibrium chiral dynamics are investigated
by using the method of the time dependent variational approach
with squeezed states as trial states. 
Our numerical simulations show that large domains of the
disoriented chiral condensate (DCC) are formed due to the combined
effect of the mode-mode and isospin-isospin correlations.
Moreover, it is found that, when the mode-mode correlation is
included, the DCC domain formation is accompanied by the amplification
of the quantum fluctuation, which implies the squeezing of the state.
However, neither the DCC domain formation nor the
amplification of the quantum fluctuation is observed if only
the isospin-isospin correlation is included.
This suggests that the mode-mode coupling plays a key role
in the DCC domain formation.
\end{abstract}


\maketitle

\section{\label{sec:level1}Introduction}

The possibility of the formation of the disoriented chiral
condensate (DCC) in relativistic heavy ion
collisions has been investigated in a number of
theoretical studies.
Non-equilibrium field dynamics has been studied also
in conjunction with the structure formation in early universe.
Within the classical approximation it has been recognized that
large domains of DCC are produced during the nonequilibrium
process in the course of the time evolution from the quench
initial condition
\cite{ref:classical01, ref:classical02}.
On the other hand, in calculations including quantum 
mechanical effects,
mainly homogeneous (translationally invariant) systems
have been studied to avoid technical difficulty
in numerically calculating
the two-point functions \cite{boyanovski01, cooper01}.
In fact, there have been several attempts to include
spatial inhomogeneity \cite{cooper02, smit, bettencourt}
and also memory effect \cite{berges,ikeda} in
the quantum mechanical time evolution of the fields.
However, the correlation between modes with different momenta
(mode-mode correlation) was not taken into account
in these works and the effect of interactions has been
included only through the mean fields.
It has not been settled whether large domain structure is
formed when quantum effects are taken into account.

Recently, it was explicitly shown that DCC domains are formed in
quantum calculation for the first time
in the case of 1+1 dimensional geometry \cite{IAT}.
In that calculation,
the mode-mode correlations are explicitly taken into account and
no translational invariance was assumed.
The importance of the mode-mode correlation is
qualitatively understood as follows.
If one does not include the mode-mode correlation, each mode is
decoupled from each other during the time evolution except for the indirect
interactions through the mean fields. The energy transfer from modes with
short wavelengths to those with long wavelengths is not effective enough
to form large correlated domains.
If one includes the mode-mode correlation, 
direct energy transfer from short wavelength modes
to long wavelength modes and as a result
the amplification of long wavelength modes
in the pion fields become possible,
which can lead to the formation of large DCC domains.

In this paper, in addition to the mode-mode correlation,
we examine the effect of the correlation between 
fields having different isospin components
(isospin-isospin correlation).
This effect on the domain formation of the DCC has not been given
particular attention to in earlier studies.
However, the isospin-isospin correlation
can arise during the nonequilibrium process
as well as the mode-mode correlation. It is important
to incorporate both correlations on an equal footing.
In addition, it is of interest to study whether the isospin-isospin
correlation affects the domain formation of DCC and, if yes,
it is then important to compare the ways the two correlations affect
the domain formation. We also make further examination
on the quantum mechanical features
of the domain formation in this paper.

The rest of the paper is organized as follows.
In Section \ref{sec:eom} we briefly review the formalism and
the equations of motion, and discuss the initial condition.
In Section \ref{sec:numerical}, we present numerical results.
Section \ref{sec:conclusion} is devoted to summary and conclusions.

\section{\label{sec:eom}Equations of Motion and Initial Condition}

We take the O(4) linear sigma model as a low energy
effective theory of QCD and apply the method of the time dependent
variational approach (TDVA) with squeezed states as trial states
as in the previous paper \cite{IAT}.
This method was originally developed by Jackiw and Kerman as 
an approximation in the functional Schr\"odinger approach
\cite{JK} and later it was shown to be equivalent to TDVA with squeezed
states by Tsue and Fujiwara \cite{TF}.
This method makes it possible to solve the time-evolution
of the order parameters (mean fields),
and the quantum fluctuations and correlations
in a self-consistent manner.

\begin{widetext}
\subsection{Equations of motion}

We denote the sigma ($\sigma({\vec x})$) and pion
$({\vec \pi}({\vec x})$) field operators as a four dimensional vector,
$\phi_a({\vec x}) = (\sigma({\vec x}), {\vec \pi}({\vec x}))$,
where $a$ runs from 0 to 3.
Then the Hamiltonian $H$ of the O(4) linear sigma model is given as
\begin{eqnarray}
 H  & = & \int \sum_{a=0}^3 \biggl \{ \frac{1}{2}{\pi}
_a({\vec x})^2 + \frac{1}{2}{\vec \nabla} {\phi}_a({\vec x})\cdot
 {\vec \nabla} {\phi}_a({\vec x})
 + \lambda \bigl[ {\phi}({\vec x})^2 - v^2 \bigr]^2
 - h {\phi}_0({\vec x}) \biggr \}d{\vec x}  \nonumber \\
    & = & \int \sum_{a=0}^3 \biggl \{ \frac{1}{2}{\pi}
_a({\vec x})^2 + \frac{1}{2}{\vec \nabla} {\phi}_a({\vec x})\cdot
 {\vec \nabla} {\phi}_a({\vec x})
 + V[ \phi ({\vec x} ) ] \biggr \}d{\vec x} ,
 \label{hamiltonian:o4}
\end{eqnarray}
where ${\phi}({\vec x})^2 = \sum_{a=0}^3 {\phi}_a({\vec x})^2$,
and $\lambda$, $v$, and $h$ are constants.
Note that ${\pi}_a({\vec x})$ is the conjugate field operator
of $\phi_a({\vec x})$ and should not be confused with the pion field
operator. We determine the three model parameters,
$\lambda$, $v$, and $h$, so that they give the pion mass
$M_{\pi} = 138$ MeV, the sigma meson mass $M_{\sigma} = 500$ MeV,
and the pion decay constant $f_{\pi} = 93$ MeV
in the one loop level 
in the ``broken symmetry'' vacuum state
as in the previous study, i.e.,
$\lambda = 3.44$, $v=110$ MeV,
and $h=(103~{\rm MeV})^3$ \cite{IAT, TKI}.

The squeezed states used as trial states have the following form:
\begin{eqnarray}
 | \Phi (t) \rangle &=&
 \exp \left\{ i \sum_{a=0}^{3} \int \left(
 D_{a}({\vec x}, t) \phi_{a}({\vec x}) - C_{a}
({\vec x}, t) \pi_{a}({\vec x})\right) d {\vec x}  \right\} \nonumber \\
 &\times&  \mathcal{N}(t)
 \exp \Biggl\{ \sum_{a, b = 0}^{3} \int 
 \phi_{a}({\vec x})
 \Bigl(\frac{1}{4}G_{a b}^{(0)-1}({\vec x}, {\vec x'}) 
 - \frac{1}{4} G_{a b}^{-1}({\vec x}, {\vec x'},t)
 + i \Pi_{a b}({\vec x}, {\vec x'}, t)  \Bigr)
 \phi_{b}({\vec x})d {\vec x} d {\vec x'}\Biggr\} | 0 \rangle.
\label{squeezedstate}
\end{eqnarray}
Here $|0 \rangle$ is the vacuum state,
and $G_{a a}^{(0)}({\vec x},{\vec y}) = \langle
0|\phi_a({\vec x})\phi_a({\vec y})|0 \rangle $ and
$G_{a b}^{(0)} ({\vec x},{\vec y}) = 0$ for $a \neq b$.
$C_{a}({\vec x},t)$ and $D_{a}({\vec x},t)$ are the c-number
mean fields of $\phi_{a}({\vec x})$ and
$\pi_a({\vec x})$, respectively.
$G_{a b}({\vec x},{\vec y},t)$ and $\Pi_{a b}({\vec x},{\vec y},t)$
are the quantum correlation functions
having isospin indices $a$ and $b$
(quantum fluctuation if both ${\vec x} = {\vec y}$ and $a=b$ hold),
and the canonical conjugate variable for $G_{a b}({\vec x},{\vec y},t)$,
respectively. Thus, the trial state is specified by
$C_{a}({\vec x},t)$, $D_{a}({\vec x},t)$,
$G_{a b}({\vec x},{\vec y},t)$, and $\Pi_{a b}({\vec x},{\vec y},t)$.
All of these variables are real.
$\mathcal{N}(t)$ is a normalization constant,
whose explicit form is not needed in the calculation of 
the time evolution of observables. Its explicit expression
is given in Ref. \cite{TF}.
When there is no isospin-isospin correlation,
the above squeezed state reduces to the direct product of the squeezed
states with single isospin labels, $a=0\sim 3$.
The squeezed state include the coherent state as a special
case. In addition, correlations can also be taken into account
with the trial states. Thus, the trial states given by
Eq. (\ref{squeezedstate})
span a wide subspace in the physical Hilbert space and
are expected to be able to describe a variety of
quantum features of the system.

The time evolution is determined by the time dependent
variational principle:
\begin{equation}
\delta\int \langle \Phi (t) | i \frac{\partial ~}{\partial t} - H
| \Phi (t) \rangle dt = 0.
\end{equation}
The equations of motion in momentum space read
\begin{eqnarray}
 \ddot{C}_{a}({\vec k},t) &=& - {\vec k}^{2} C_{a}({\vec k},t)
- \mathcal{M}_{a}^{(1)}({\vec k},t), \nonumber \\
\dot{G}_{a b}({\vec k},{\vec k'},t) &=& 2 \sum_{a' = 0}^{3} \langle
 {\vec k} | \Bigl[
 G_{a a'}(t) \Pi_{a' b}(t)
+ \Pi_{a a'}(t) G_{a' b}(t) \Bigr]
 | {\vec k'} \rangle , \nonumber \\ 
\dot{\Pi}_{a b}({\vec k},{\vec k'},t)&=& \frac{1}{8}
 \sum_{a' = 0}^{3} \langle {\vec k} | G_{a a'}^{-1}(t) G_{a' b}^{-1}(t)
 | {\vec k'} \rangle 
- 2 \sum_{a' = 0}^{3} \langle {\vec k} | \Pi_{a a'}(t) \Pi_{a' b}(t)
 | {\vec k'} \rangle 
- \frac{1}{2} (2 \pi)^3 {\vec k}^{2} \delta_{a b} \delta ^{3} ( {\vec k}
- {\vec k'} )
 - \frac{1}{2} \mathcal{M}_{a b}^{(2)} ({\vec k}-{\vec k'},t),
 \nonumber \\
\mathcal{M}_{a}^{(1)}({\vec k},t) &=&
\Bigl[ - m^{2} C_{a}({\vec k},t) 
+ \int \frac{d {\vec l}d {\vec l'}}{(2 \pi)^{6}}
\Big( 4 \lambda C_{a} ({\vec k}- {\vec l} - {\vec l'},t)
\sum_{b = 0}^{3} C_{b}({\vec l},t) C_{b}({\vec l'},t) 
+ 12 \lambda C_{a}({\vec k} - {\vec l} - {\vec l'},t)
G_{a a}({\vec l},{\vec l'},t) 
\nonumber \\
& & + 4 \lambda C_{a}({\vec k} - {\vec l} - {\vec l'},t) \sum_{b (\neq a)} 
G_{b b} ({\vec l},{\vec l'},t)
+ 8 \lambda \sum_{b (\neq a)} C_{b}({\vec k} - {\vec l} - {\vec l'},t) 
G_{b a} ({\vec l},{\vec l'},t) \Bigr) - h \delta_{a0} V \Bigr] , \nonumber \\
\mathcal{M}_{a a}^{(2)}({\vec k},t) &=&
 - m^{2} V + \int \frac{d {\vec l}} {(2 \pi )^{3} } \Bigl( 12 \lambda \bigl( 
 C_{a}({\vec k} - {\vec l},t) C_{a}({\vec l},t) 
 + G_{a a}({\vec k} - {\vec l},{\vec l},t) \bigr) \nonumber \\
 & & + 4 \lambda \sum_{b (\neq a)}
\bigl( C_{b}({\vec k} - {\vec l},t) C_{b}({\vec l},t)
 + G_{b b}({\vec k} - {\vec l},{\vec l},t) \bigr)\Bigr), 
\nonumber \\
\mathcal{M}_{a b(\neq a)}^{(2)}({\vec k},t) &=&
 8 \lambda \int \frac{d {\vec l}} {(2 \pi )^{3}}
\Bigl( C_{a}({\vec k} - {\vec l},t) C_{b}({\vec l},t)
 + G_{a b}({\vec k} - {\vec l},{\vec l},t) \Bigr)
\label{eom},
\end{eqnarray}
\end{widetext}
where $m^2  =  4 \lambda v^2$ and $V={\rm Tr}{\bf 1} = \int d {\vec x}$.
$C_a({\vec k},t)$ is
the mean field for the $\phi_a$ field with momentum ${\vec k}$, and
$G_{a b}({\vec k},{\vec k'},t)$ and $\Pi_{a b}({\vec k},{\vec k'},t)$ are
the correlation between modes with isospins $a$ and $b$, and
momenta ${\vec k}$ and ${\vec k'}$
(the quantum fluctuation for $a = b$ and ${\vec k} = {\vec k'}$), and
the canonical conjugate variable for $G_{ab}({\vec k},{\vec k'},t)$,
respectively.
The one-point functions $C_a({\vec k},t)$ and $D_a({\vec k}, t)$
and the two-point functions
$G_{ab}({\vec k}, {\vec k'}, t)$ and $\Pi_{ab}({\vec k}, {\vec k'}, t)$
are the momentum space representations of
$C_a({\vec x},t)$, $D_a({\vec x}, t)$, 
$G_{ab}({\vec x}, {\vec x'}, t)$, and $\Pi_{ab}({\vec x}, {\vec x'}, t)$,
respectively. In Eq. (\ref{eom}), we have used the following notation,
\begin{equation}
\langle {\vec k} | H(t) I(t) | {\vec k'} \rangle =
  \frac{1}{(2\pi)^{3}} \int
  H({\vec k},{\vec k''},t) I({\vec k''},{\vec k'},t) {d {\vec k''}}.
\end{equation}

$\mathcal{M}_{a b}^{(2)}({\vec k}, t)$ in Eq. (\ref{eom}) originates
from the nonlinear coupling term in the Hamiltonian and is given as
follows:
\begin{eqnarray}
\mathcal{M}_{a b}^{(2)}({\vec k}, t)
& = & \int \mathcal{M}_{a b}^{(2)}({\vec x}, t)
e^{i{\vec k} \cdot {\vec x}} {d {\vec x}} , \nonumber \\
\mathcal{M}_{a b}^{(2)}({\vec x}, t) & = &
\exp \left \{ \frac{1}{2} \sum_{a',b' =0}^{3}
\frac{\partial}{\partial z_{a'}}G_{a' b'}({\vec x},{\vec x},t)
\frac{\partial}{\partial z_{b'}} \right \} \nonumber \\
& & \left . \times \frac{\partial^2 }{\partial z_a \partial z_b} U[z]
\right | _{z_{a}=C_{a}({\vec x},t)} ,
\end{eqnarray}
where $U[z]$ represents the c-number potential that
is obtained by replacing $\phi_a$ in the potential term of
the model Hamiltonian $V[\phi]$ by $z_a \in {\bf R} $:
\begin{equation}
U [z]  =  \left . \phantom{\frac{}{}} V [\phi ]
\right | _{\phi_a ({\vec x}) \, =\, z_a} \quad (a\, = \, 0-3).
\end{equation}
The third equation in Eq. (\ref{eom})
tells us that mode-mode correlations
in momentum space arise through
$\mathcal{M}_{a b}^{(2)}({\vec k} - {\vec k'}, t)$
with $\vec k \neq \vec k'$ even if there is initially no such correlation.
Similarly, the isospin-isospin correlation
arise due to the interaction term
$\mathcal{M}^{(2)}_{a b}({\vec k} - {\vec k'}, t)$ with $a \neq b$.

In the numerical calculation, we assume that the system is in a cubic
box with the volume $V = L^3$ and complies the periodic boundary condition,
and that the lattice spacing in each direction is the same,
$d = \frac{L}{N}$, where $N$ is the division number in each direction.
Accordingly, the momentum $\vec{k}$,
the square momentum $\vec{k}^2$, and the momentum integral
in Eq. (\ref{eom}) are replaced as follows:
\begin{eqnarray}
{\vec k} & \rightarrow & \frac{2 \pi}{L} {\vec n}, \nonumber \\
{\vec k}^2 & \rightarrow & \sum_{i = x,y,z} \frac{4}{d^2}
\sin^2 \left ( \frac{\pi n_i}{N} \right ), \nonumber \\
\int \frac{d {\vec k}}{(2 \pi)^3} & \rightarrow & \frac{1}{V}
\sum_{n_x =0}^{N-1} \sum_{n_y =0}^{N-1} \sum_{n_z =0}^{N-1},
\end{eqnarray}
with $\vec{n} = (n_{x},n_{y},n_{z})$.

\subsection{Initial condition}
\label{subsec_initcond}

As the initial condition, we take the quench initial
condition.
In the quench scenario, the following assumption on the
time evolution of the system is made:
The matter produced in relativistic heavy-ion collisions
is initially thermalized and that chiral symmetry
is restored. Then, this matter goes through
rapid cool down and the true vacuum spontaneously
breaks chiral symmetry. 
In this process, however, the change of the effective
potential is so rapid that the order parameter cannot follow
the change and remains around the origin, where the minimum of
the effective potential is in the chirally symmetric phase.
\cite{ref:classical01, ref:classical02, IAT}. 
Namely, in the quench scenario, the mean of the chiral fields as 
well as the correlation of the fields are the same as those
before the sudden cooling (quenching), but only the sizes of
the fluctuations of the mean fields and their conjugate fields
are quenched.
In order to realize the quench scenario in our numerical
calculation, we specify the initial condition
for the mean fields and the quantum correlation and fluctuation
as follows.

For the mean fields, we have used the same prescription for the
initial condition as in Refs. \cite{IAT, AM}:
at each grid site, the mean field variables
for the chiral fields and their conjugate variables
$C_a({\vec x},0)$ and $D_a({\vec x},0)$ are randomly distributed
with the Gaussian weights with the following parameters,
\begin{eqnarray}
 \langle C_a({\vec x},0) \rangle &=& 0, \nonumber \\
 \langle C_a({\vec x}, 0)^2 \rangle
 - \langle C_a({\vec x}, 0) \rangle ^2 &=& \delta^2 , \nonumber \\
 \langle D_a({\vec x}, 0) \rangle &=& 0 , \nonumber \\
 \langle D_a({\vec x}, 0)^2 \rangle
 - \langle D_a({\vec x},0) \rangle ^2 &=& \frac{D}{d^2} \delta ^2 ,
 \label{initial:mf4}
\end{eqnarray}
where $\delta$ is the Gaussian width
and $D$ is the spatial dimension.
In relating the Gaussian widths of $C_a({\vec x}, 0)$ 
and $D_a({\vec x}, 0)$,
we have taken advantage of the virial theorem.
For detailed discussion on the use of virial theorem,
see Ref.~\cite{AM}.
We use the Gaussian width $\delta = 0.19 v$,
which is the same value which has been taken in
Ref. \cite{IAT}. This value is so chosen that the
fluctuation is small enough and the chiral symmetry is
spontaneously broken. 
This is necessary to ensure that the quench scenario
is simulated. This value of $\delta$ is taken as a typical value
and, of course, other choices are also possible.

In order to fix the initial condition for the quantum
correlation and fluctuation,
we have assumed that each of the sigma and pion states
in momentum space is independently in a coherent state
with a degenerate mass $m_{0}$. Accordingly,
$G_{a b} ({\vec x}, {\vec y}, 0)$ and $\Pi_{a b}({\vec x}, {\vec y}, 0)$
are given as follows:
\begin{eqnarray}
 G_{a a}({\vec x}, {\vec y},0) &=&
 \frac{1}{(2 \pi)^{3}}\int_{0}^{\Lambda} 
 \frac{e ^{i {\vec k}\cdot({\vec x}- {\vec y})}} {2 \omega_k}
 d {\vec k}, \nonumber  \\
 G_{a b(\neq a)} ({\vec x}, {\vec y}, 0) &=& 0 , \nonumber \\
 \Pi_{a b}({\vec x}, {\vec y}, 0) &=& 0 , \label{coh_fluc}
\end{eqnarray}
with $\omega _{k} = \sqrt{m_{0}^2 + {\vec k}^2}$.
We adopt $m_{0} = 200$ MeV.
The two-point functions $G_{a b}({\vec x},{\vec y},0)$
and $\Pi_{a b}({\vec x},{\vec y},0)$ are thus diagonal in both
momentum and isospin spaces in the initial state. 
However, as previously explained, their off-diagonal components
in both momentum and isospin spaces arise through the non-linear
coupling term in the equation of motion Eq. (\ref{eom}), if the mean
field is not translationally invariant in coordinate space
or symmetric in isospin space.
We examine the effect of these off-diagonal components of the
Green's function on the time evolution of the system in the next
section.

\begin{figure*}[!tbh]
  \resizebox{160mm}{!}{\includegraphics[angle=0]{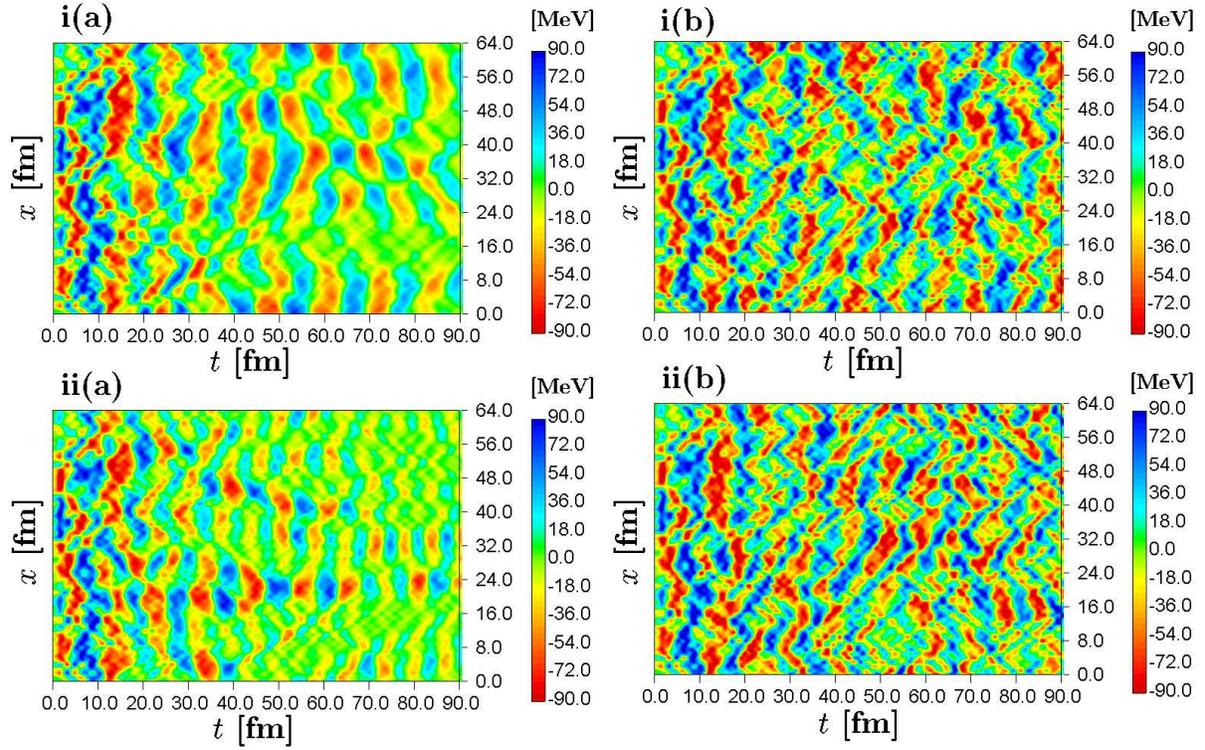}}
  \caption{(Color online) Time evolution of
   the mean field of the third component of
   the pion field in cases i(a), i(b), ii(a), and ii(b). }
  \label{meanfield}
\end{figure*}

\begin{figure*}[!thb]
\begin{minipage}{.45\linewidth}
\includegraphics[width=60mm]{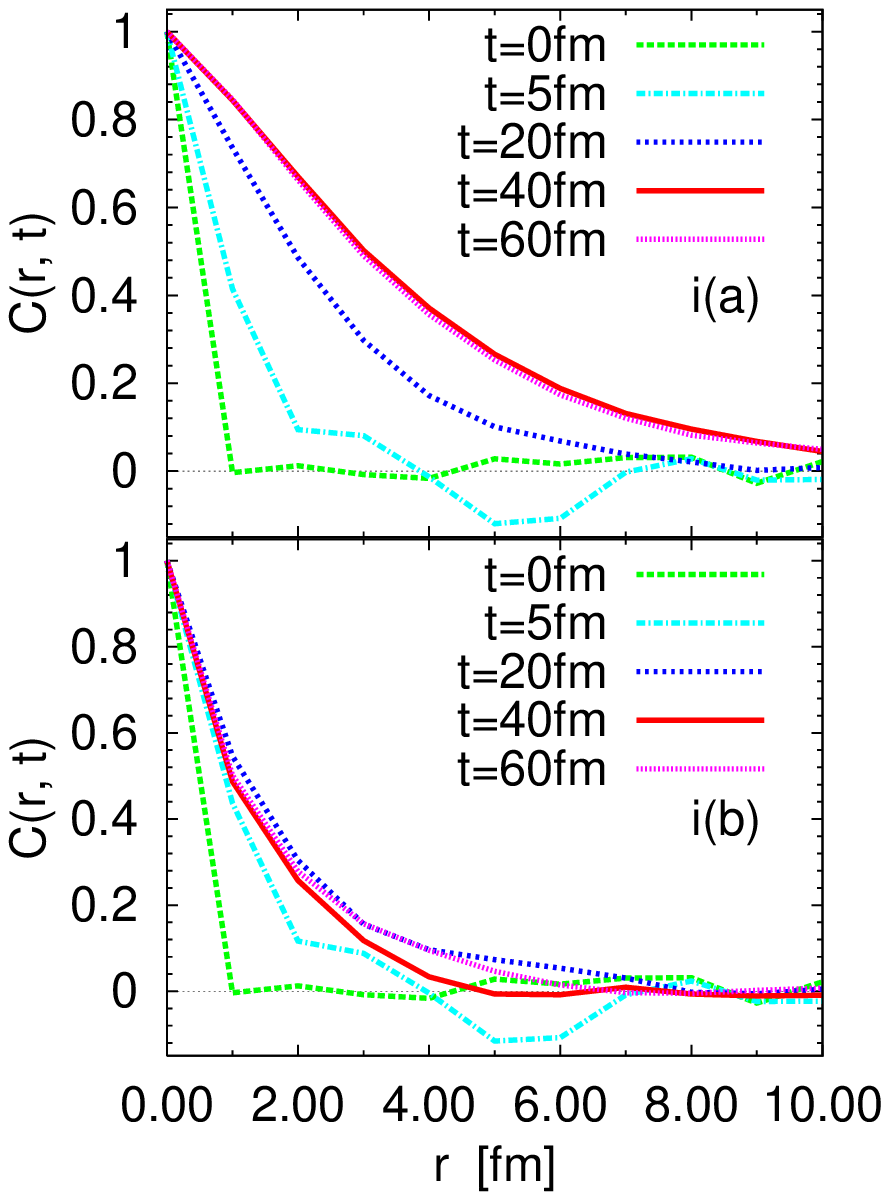} 
              \caption{(Color online) 
                 Snap shots of the
                 spatial correlations of the mean fields
                 in case i(a) and case i(b). }
               \label{correlation1}
\end{minipage}
\hspace{5mm}
\begin{minipage}{.45\linewidth}
\includegraphics[width=60mm]{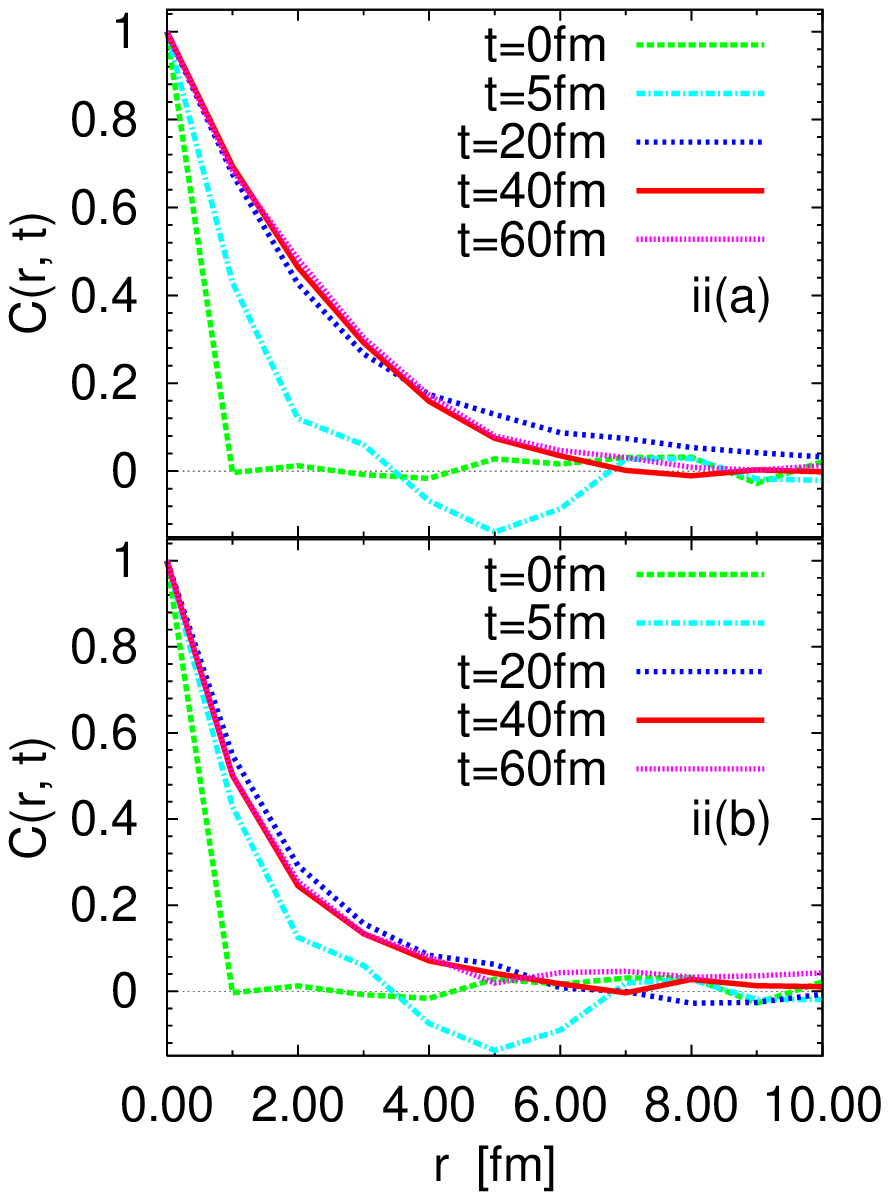} 
               \caption{(Color online) 
                 Snap shots of the
                 spatial correlations of the mean fields
                 in case ii(a) and case ii(b).}
               \label{correlation2}
\end{minipage}
\end{figure*}

\section{\label{sec:numerical} Numerical Results}

We evolve the mean fields and the two-point functions
on a discrete lattice with the total length in the $z$ direction
$L = 64$ fm and the lattice spacing $d = 1.0$ fm,
which corresponds to the three dimensional
momentum cutoff $\Lambda = 1071$ MeV.
We assume translational invariance in the
$x$ and $y$ directions, thus $D=1$ in Eq. (\ref{initial:mf4}).
We have confirmed that the total energy is conserved
at least to an accuracy of less than 1 percent
throughout the time evolution.

To understand the role of
the mode-mode and isospin-isospin correlations separately,
we have performed the following four numerical simulations:\\
\noindent Case i(a): The mode-mode and the isospin-isospin
correlations are both included.\\
\noindent Case i(b): The mode-mode correlation is ignored, but
the isospin-isospin correlation is included.\\
\noindent Case ii(a): The mode-mode correlation is
included, but the isospin-isospin correlation is ignored.\\
\noindent Case ii(b): The mode-mode and the isospin-isospin
correlations are both ignored.\\

\subsection{Time evolution of the mean fields}\label{time-evolution_mf}

First we show the time evolution of the mean fields.
Figure \ref{meanfield} illustrates the
time evolution of the mean field of the third component of the pion
field in the four cases.
The value of the mean field
is represented by different colors at each position and time.

In case i(a) and case ii(a),
where the mode-mode correlation is included,
the formation of the domain structure of the pion field
is clearly seen. Also it is observed that
the size of the domains continues to grow beyond the time scale
of the rolling down, $\sim$a few fm, in these cases.
On the contrary, the formation of only small domains is
observed in case i(b) and case ii(b).
In these cases, the mode-mode correlation is not included
and short range fluctuation is dominant throughout the time
evolution. No qualitative change in the behavior
of the mean fields is observed after a few fm.
These results imply that the mode-mode correlation is a key
ingredient for the DCC domain formation.

Furthermore, it is observed that the domain structure
is larger in case i(a),
where the mode-mode and isospin-isospin correlations
are both included, compared with case ii(a),
where only the mode-mode correlation is included.
The appearance of the larger domains in case i(a)
is due to the combined effect of
mode-mode and isospin-isospin correlations.
Also it is observed that in case i(b),
where the isospin-isospin correlation is included but the
mode-mode correlation is not included,
domains are slightly larger compared to those in
case ii(b), where either correlation is not included.
From these results, we find that
the domain formation is most effective when both 
mode-mode and isospin-isospin correlations are included.

\subsection{Spatial Correlation}\label{spatial_correlation}
In this subsection, we show the spatial correlation function
$C(r, t)$ defined by
\begin{equation}
 C(r, t) = \frac{\int 
           {{\vec C} }({\vec x}) \!\cdot\! {{\vec C}}({\vec y})
           \delta (|{\vec x} - {\vec y}| - r) d{\vec x} d{\vec y} }
           {\int | {{\vec C}}({\vec x}) | | {{\vec C}}({\vec y}) |
           \delta (|{\vec x} - {\vec y}| - r) d{\vec x} d{\vec y} }~,
\end{equation}
where ${{\vec C} }({\vec x})\!\cdot\! {{\vec C}}({\vec y})
= \sum_{i=1}^{3} {C_i}({\vec x}) {C_i}({\vec y})$
and $|{{\vec C}}({\vec x}) | =
\sqrt{ \sum_{i=1}^{3} {C_i}({\vec x})^2 }$.
This is a measure for the closeness of the order parameters
in isospin space at distance $r$ in coordinate space.

The snap shots of the spatial correlations at $t=0$
(the initial time), 5, 20, 40, and 60 fm are shown 
in Figs.~\ref{correlation1} and \ref{correlation2} for
the four cases.
The correlation functions are 
obtained by taking the ensemble average over 10 independent
initial states.
At the initial time $t = 0$,
the spatial correlation length is almost
equal to the lattice spacing in each case.
This is so, because the initial mean fields are independently
distributed at each grid site.
However, difference in the correlation
function appears as the time elapses.

The correlation lengths in the upper figures
are always longer than those in the lower figures
in both Fig.~\ref{correlation1} and Fig.~\ref{correlation2}.
This tells us the importance of the mode-mode correlation
in the trial quantum states in the growth of the correlation length
regardless of the presence of the isospin-isospin
correlation.
Moreover, the correlation length in case i(a) is
longer than in case ii(a), while that in case
i(b) is only a little longer than in case ii(b).
The formation of larger DCC domains is induced
by the combined effect of the mode-mode correlation and
the isospin-isospin correlation. 

The duration of the correlation generation
differs between cases i(a) and i(b), and between cases ii(a) and ii(b).
However, it is almost the same in cases i(a) and ii(a), and in cases
i(b) and ii(b).
In cases i(b) and ii(b),
the formation of the correlation almost finishes by $t=5$ fm,
which is the typical time scale within which the rolling-down of the
order parameter from the top of the effective potential
is completed \cite{gm}. 
We will explicitly show this in the next subsection.
The fact that the correlation increases
beyond that time scale in cases i(a) and ii(a)
strongly suggests that the mode-mode coupling is another
driving force for the domain formation in addition to
the instability of the low momentum modes that exists
during the rolling-down of the order parameter \cite{ref:classical01}.

As a possible mechanism of DCC formation, the parametric
resonance has been also proposed \cite{ref:parametric}.
It is, however, also understood within the framework of
the mean field theory without correlations among modes;
it is in essence a one-mode problem.
Thus, the mode-mode coupling is yet another mechanism for DCC formation,
and according to the result of our calculation, it has the
most dominant effect on the generation of the field correlations.
It is interesting to compare this result with the result
of the classical simulation by
Rajagopal and Wilczek \cite{ref:classical01}.
In their work, it was found that
the amplification of low momentum modes lasts far beyond the typical
time scale of the rolling-down of the order parameter. Their calculation
is classical, but includes direct couplings among modes.
This also exemplifies the importance of the direct mode-mode coupling
in the domain formation.

\begin{figure}[t]
  \resizebox{85mm}{!}{\includegraphics[angle=0]{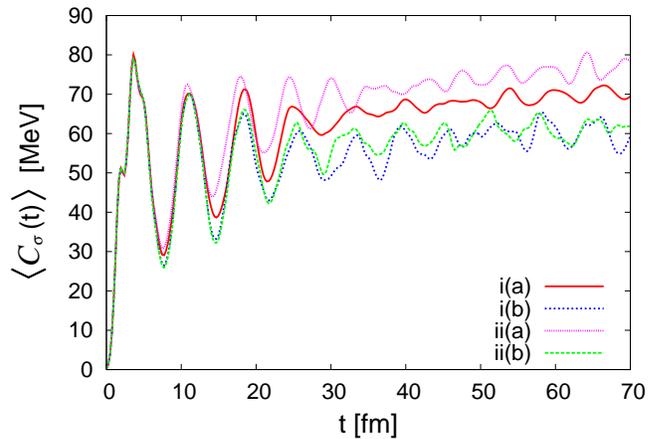}}
  \caption{(Color online) Time evolution of
   the spatial average of the sigma field
  in cases i(a), i(b), ii(a), and ii(b).}
  \label{sigma_evolution}
\end{figure}

\subsection{Time evolution of the sigma field}
In this subsection, 
we examine whether there is the collective oscillation
of the sigma field, which is required for the 
DCC domain formation through the parametric resonance.
For this purpose, we show the time evolution of the spatial 
average of the sigma field defined by,
\begin{equation}
 \langle C_{0} (t) \rangle = \frac{1}
{V} \int C_{0} ({\vec x},t) d {\vec x}.
\label{average_sigma}
\end{equation}
Note that the zeroth component of the chiral field is
the sigma field.

Figure \ref{sigma_evolution} shows the time evolution of
the spatial average of the sigma field
in the four cases. 
This result was obtained by taking the ensemble average
over 10 initial states.
In each case the sigma field quickly rolls down from the top
of the effective potential and approaches its ground state
value within less than 5 fm. After that, the sigma field
continues to oscillate as found in earlier studies
\cite{ref:classical01,ref:classical02}.
The oscillation is damped more quickly in cases i(a) and ii(a),
where the direct mode-mode coupling is taken into account.
In case i(a), the oscillation almost disappears by $t=25$ fm.
On the other hand, according to Fig.~\ref{correlation1},
the correlation length of the pion field continues to grow
until about $t=40$ fm. Thus, the collective oscillation of
the sigma field is not necessarily needed for the DCC
domain formation unlike in the parametric resonance scenario.

\begin{figure}[t]
\includegraphics[width=85mm]{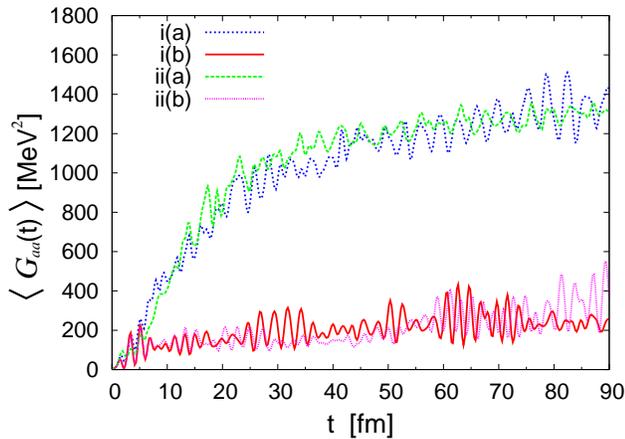}
                \caption{(Color online) Time evolution of
the spatial average of the quantum fluctuation of
the third component of the pion field,
$\langle G_{33}(t) \rangle$ in cases i(a), i(b), ii(a), and ii(b). }
\label{fluctuation}
\end{figure}

\subsection{Quantum fluctuations}
In this subsection,
we show the time evolution of the quantum fluctuation.

Figure \ref{fluctuation} depicts the time evolution of
the spatial average of the quantum fluctuation of the third
component of the pion field, $\langle G_{33}(t)\rangle$,
which is defined by,
\begin{equation}
  \langle G_{33}(t) \rangle = \frac{1}{V}
  \int G_{33}({\vec x} , {\vec x}, t) d{{\vec x}} .
\end{equation}
In this figure, no event average is taken.
However, since the spatial size of the system is large
enough, the behavior of $\langle G_{11}(t) \rangle$ and
$\langle G_{22}(t) \rangle $ is expected to be similar
to that of $\langle G_{33}(t) \rangle$ due to the isospin
symmetry.

In the initial state, fluctuation is small.
This is because the initial state is chosen to be
the direct product of the coherent states.
The remarkable feature is that 
the quantum fluctuation is substantially amplified
in the cases with the mode-mode correlation,
i.e., cases i(a) and ii(a). 
On the other hand, such substantial amplification of the quantum
fluctuation is absent in the cases without mode-mode correlation,
i.e., cases i(b) and ii(b).
In addition, it is noted that
the duration of the amplification of the quantum
fluctuation in cases i(a) and i(b) is of the same
order as that of the correlation formation observed
in Fig.~\ref{correlation1}. 
In the space of the trial states,
the amplification of the quantum fluctuation
implies squeezing of the states.
Thus, the main part of the late time DCC domain formation occurs
in concurrence with squeezing of the states.

On the contrary, only little difference in the time evolution
of the quantum fluctuation is observed between cases
i(a) and i(b), and also between cases ii(a) and ii(b).
This implies that 
the isospin-isospin correlation does not lead to
further amplification of the quantum fluctuation.

However, we have observed that there is a little difference
in the formation of correlations between cases i(a) and i(b), and
also between cases ii(a) and ii(b), in Figs. \ref{correlation1} and
\ref{correlation2}. This suggests that there are at least
two mechanisms that lead to the long range correlation of the
pion fields: 
one that is accompanied by
the squeezing of the pion fields
and the other that is not.

\begin{figure}[t]
\includegraphics[width=85mm]{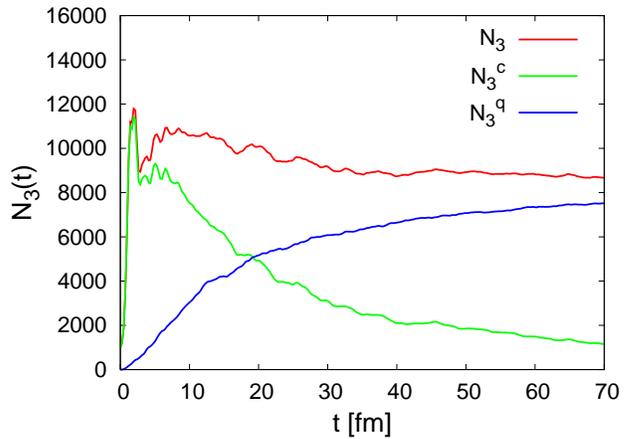}
	\caption{(Color online) Time evolution of the total particle numbers of
	the third component of the pion field $N_{3}$,
        its classical part $N_{3}^{cl}$, and its quantum part
	$N_{3}^{q}$ in case ii(a).}
        \label{snumber1}
\end{figure}

\subsection{Pion particle numbers}

In this subsection, we calculate the particle numbers associated with
the pion fields for cases ii(a) and ii(b).

In order to analyze the particle numbers associated with the pion fields,
we expand the field operator $\phi_a({\vec x})$ and its
conjugate operator $\pi_a({\vec x})$ by the
annihilation operator ${\hat a}_{a}({\vec k})$
and creation operator ${\hat a}_{a}^{\dagger}({\vec k})$
of the pion with the momentum ${\vec k}$ and isospin $a$,
\begin{eqnarray}
\phi_{a}({\vec x}) &=& \frac{1}{(2 \pi)^3}
\int \sqrt{\frac{1}{2
\omega_k}} \left[ {\hat a}_{a}({\vec k}) e^{i {\vec k} \cdot {\vec x}}
+ {\hat a}_{a}^{\dagger}({\vec k})
e^{- i {\vec k} \cdot {\vec x}} \right] {d {\vec k}}, \nonumber \\
\pi_a ({\vec x}) &=& \frac{1}{i (2 \pi)^3}
\int \sqrt{\frac{
\omega_k}{2}} \left[ {\hat a}_{a}({\vec k}) e^{i {\vec k} \cdot {\vec x}}
- {\hat a}_{a}^{\dagger}({\vec k})
e^{- i {\vec k} \cdot {\vec x}} \right] {d {\vec k}}, \nonumber
\end{eqnarray}
where $\omega_k$ is the energy of the normal modes with the pion
mass in the vacuum $M_{\pi}$, $\omega_k = \sqrt{M_{\pi}^2 + {\vec k}^2}$.
From these relations, we obtain
\begin{eqnarray}
 {\hat a}_{a}({\vec k}) &=& \int  \left[
 \sqrt{\frac{\omega_k}{2 }} \phi_a({\vec x})
 + i \sqrt{\frac{1}{2  \omega_k}} \pi_{a}({\vec x})
\right] e^{-i {\vec k} \cdot {\vec x}} d {\vec x}, \nonumber \\
 {\hat a}_{a}^{\dagger}({\vec k}) &=& \int \left[
 \sqrt{\frac{\omega_k}{2 }} \phi_a({\vec x})
 - i \sqrt{\frac{1}{2  \omega_k}} \pi_{a}({\vec x})
\right] e^{i {\vec k} \cdot {\vec x}} d {\vec x}.
\nonumber
\end{eqnarray}

\begin{figure}[t]
\includegraphics[width=85mm]{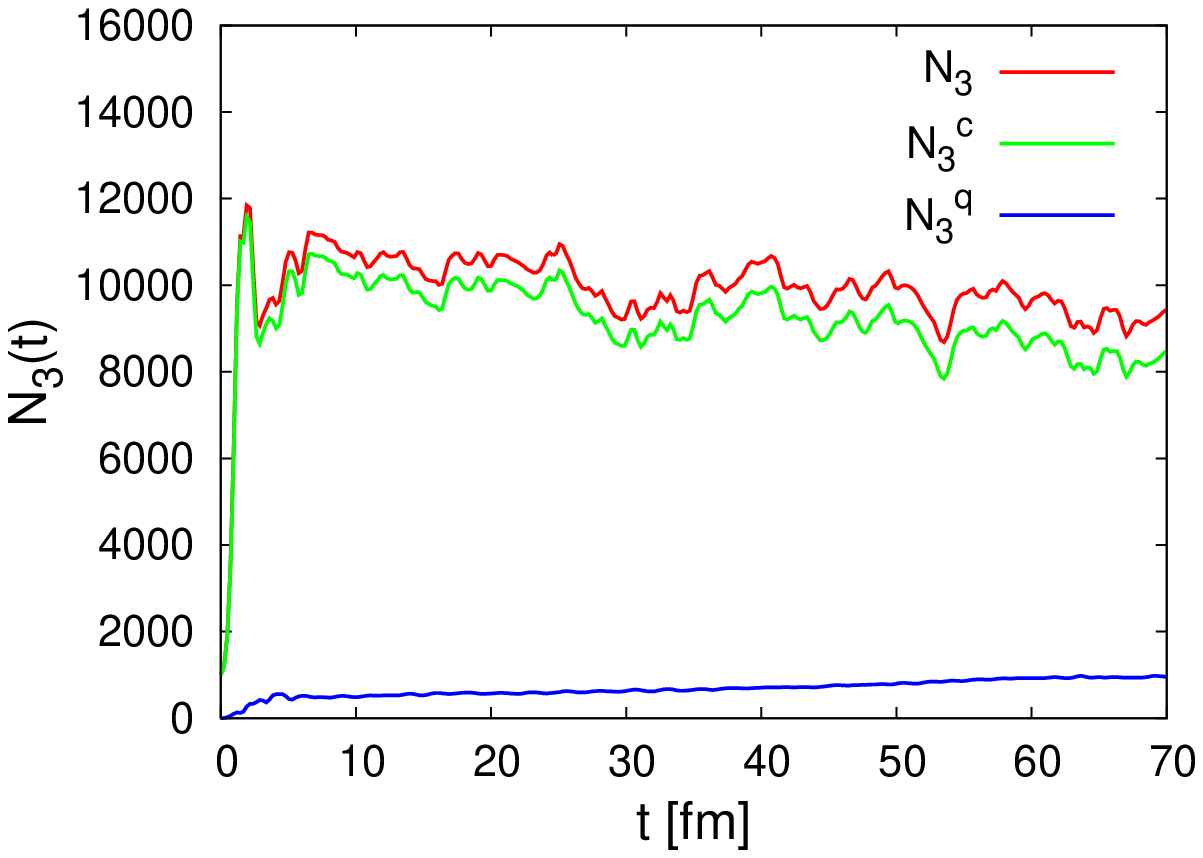}
	\caption{(Color online) Time evolution of the total particle numbers of
	the third component of pion field $N_{3}$,
        its classical part $N_{3}^{cl}$, and its quantum part
	$N_{3}^{q}$ in case ii(b).}
        \label{snumber2}
\end{figure}

Then, the expectation value of the
particle number operator of the pion with isospin $a$
in the squeezed state is given as follows:
\begin{eqnarray}
 n_{a}({\vec k}, t)  &=&
\langle \Phi (t) | 
{\hat a}_{a}^{\dagger}({\vec k}) {\hat a}_{a}({\vec k})
 | \Phi (t) \rangle  \nonumber \\
 &=& n_{a}^{cl}({\vec k}, t) + n_{a}^{q}({\vec k}, t), \nonumber \\
 n_{a}^{cl}({\vec k}, t) &=& 
 \frac{\omega_k}{2} C_{a}^{*}({\vec k}, t)C_{a}({\vec k}, t)
 + \frac{1}{2 \omega_k} D_{a}^{*}({\vec k}, t) D_{a}({\vec k}, t)
 \nonumber \\
 &-& \frac{i}{2} \bigl( C_{a}^{*}({\vec k}, t) D_{a}({\vec k}, t)
 - C_{a}({\vec k}, t) D_{a}^{*}({\vec k}, t) \bigr),
 \nonumber \\
 n_{a}^{q}({\vec k}, t) &=& 
 \frac{\omega_k}{2} G_{a a}({\vec k}, {\vec k}, t) \nonumber \\
 &+& \frac{1}{2 \omega_k} \bigl(
 \frac{1}{4} G_{a a}^{-1}({\vec k}, {\vec k}, t)
 + 4 \langle {\vec k} | \Pi_{a a} G_{a a} \Pi_{a a} | {\vec k} \rangle
 \bigr)
 \nonumber \\
 &+& i \langle {\vec k} | ( G_{a a} \Pi_{a a} - \Pi_{a a}
 G_{a a} ) | {\vec k} \rangle ,
\label{numberdensities}
\end{eqnarray}
where we have divided $n_{a}({\vec k}, t)$ into
the classical part $n_{a}^{cl}({\vec k}, t)$
and the quantum part $n_{a}^{q}({\vec k}, t)$;
$n_{a}^{cl}({\vec k}, t)$ includes the mean field and its time derivative,
$C_a({\vec k},t)$ and $D_a ({\vec k},t)$, and $n_{a}^{q}({\vec k}, t)$ includes
the quantities associated with the quantum fluctuation.
On the right hand side of Eq. (\ref{numberdensities}),
no sum over $a$ is taken. In particular,
no off-diagonal quantity with regard to the isospin indices
appear on the right hand side of the last equation of
(\ref{numberdensities}) because no isospin-isospin correlation
exists in either case ii(a) or case ii(b).
In addition, we define the total particle numbers
as the sum of the particle numbers over all modes,
\begin{eqnarray}
 N_{a}(t) &=& \sum_{{\vec k}} n_{a}({\vec k}, t) ,\nonumber \\
 N_{a}^{cl}(t) &=& \sum_{{\vec k}} n_{a}^{cl}({\vec k}, t) ,\nonumber \\
 N_{a}^{q}(t) &=&  \sum_{{\vec k}} n_{a}^{q}({\vec k}, t).
\label{sum_of_number}
\end{eqnarray}

We have calculated these particle numbers
by taking the ensemble average
over 10 different initial field configurations.
Figs.~\ref{snumber1} and \ref{snumber2} show the time
evolution of $N_{3}(t)$, $N_{3}^{cl}(t)$,
and $N_{3}^{q}(t)$ in cases ii(a) and ii(b), respectively.

\begin{figure}[t]
         \resizebox{90mm}{!}{\includegraphics{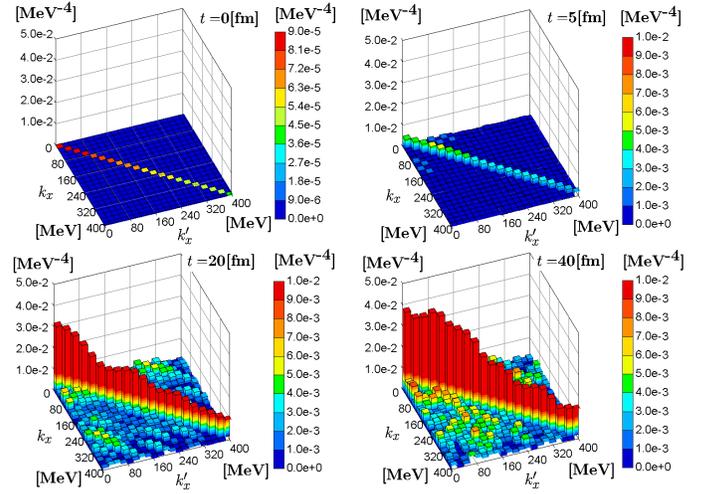}}
         \caption{(Color online) Snap shots of the two-point function in momentum space,
         $|G_{33}({\vec k},{\vec k'},t)|$, in case i(a).
         Note that the scale of the color legend at the initial time $t=0$
	 is different from the others. }
         \label{greens_snap_shot}
\end{figure}

We observe that the total
particle number $N_{3}(t)$ rapidly increases within the
typical time scale of the rolling-down of the order parameters
(a few fm) in both cases.
At first, the classical part $N_{3}^{cl}(t)$ is
dominant in both cases. In case ii(a), where 
the mode-mode correlation is included, the classical
part $N_{3}^{cl}(t)$ begins to decrease and the quantum part 
$N_{3}^{q}(t)$ begins to increase after a few fm, while the sum remains
almost unchanged. The typical time scale of the increase
of the quantum part is again close to that of the
formation of the long range correlation.
This clear separation of the stages in the time evolution
of $N_{3}^{cl}(t)$ and $N_{3}^{q}(t)$ also shows that the main
mechanism for the domain formation is not the rolling-down of
the sigma field and the associated instability of pion fields
with low momenta.
On the other hand, in case ii(b), where less correlation
results, the classical part of the total particle number
$N_{3}^{cl}(t)$ is dominant throughout the time evolution.

\begin{figure}[t]
         \resizebox{90mm}{!}{\includegraphics{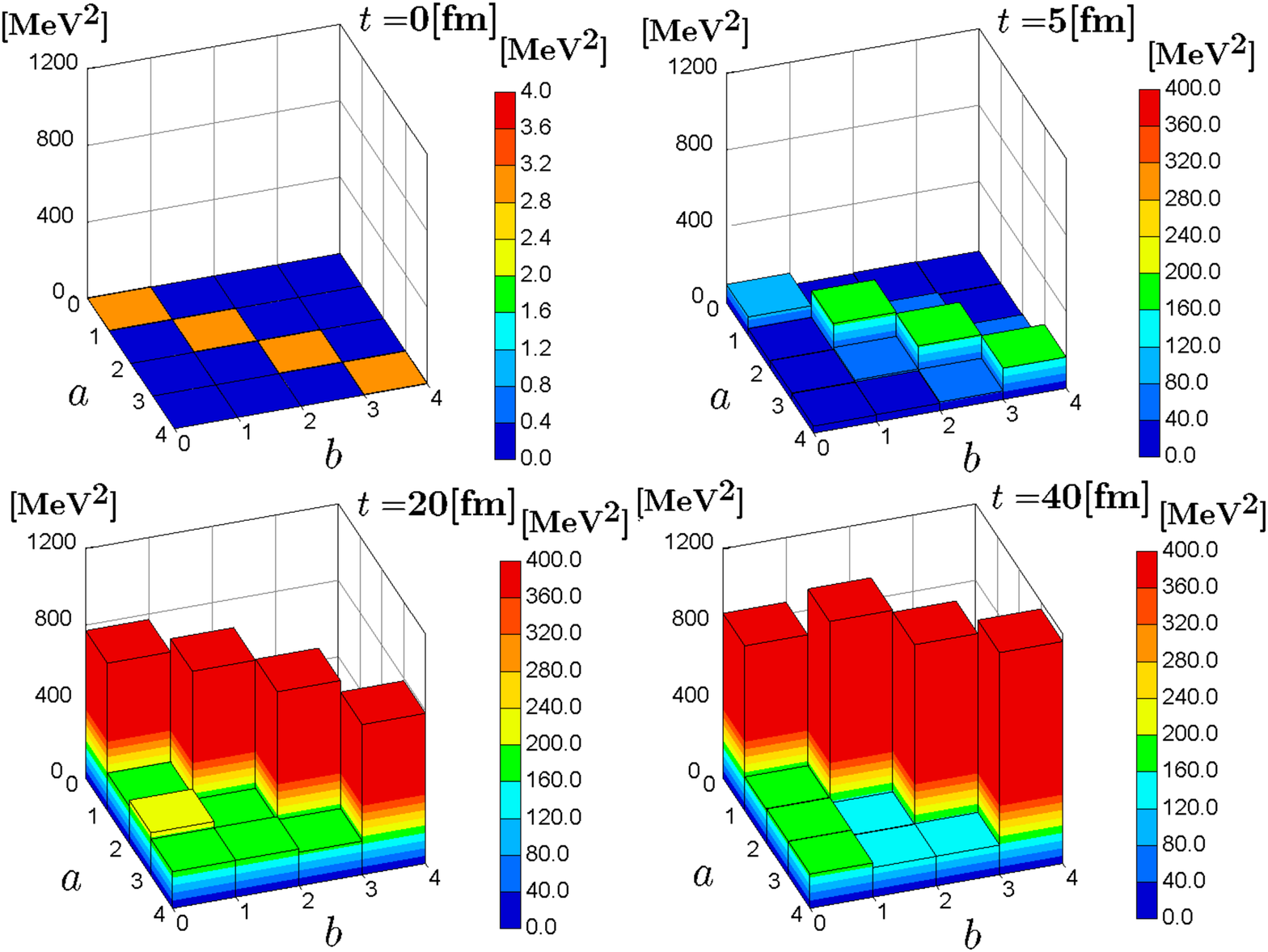}}
         \caption{(Color online) Snap shots of the two-point function
	 in isospin space, $\langle |G_{a b} (t)| \rangle$, in case i(a).
         Note that the scale of the color legend at the initial time $t=0$
	 is different from the others.}
         \label{isospin_off-diagonal}
\end{figure}

\subsection{Off-diagonal components of the two-point functions
in momentum space and isospin space}

In this subsection, we 
show the time evolution of
the off-diagonal components
of the two-point functions in momentum space and also in isospin space,
which correspond to the mode-mode correlation and
isospin-isospin correlation, respectively.

Figure \ref{greens_snap_shot} shows the absolute value of
the two-point function within the third component of the
pion field in momentum space, $|G_{33}({\vec k},{\vec k'},t)|$, 
in case i(a). In Fig. \ref{greens_snap_shot}, no event average is
taken. As explained in Section \ref{sec:eom},
in general cases where
there is no translational invariance,
the off-diagonal components of the two-point functions
develop and grow as the system evolves in time
even if they are initially absent.
According to Fig.~\ref{greens_snap_shot},
the off-diagonal components indeed appear as time elapses.
Although the diagonal components are always dominant,
the off-diagonal components are not negligible, in particular,
at low momenta. This existence of the off-diagonal components is
crucial for the DCC formation and the evolution of the
quantum fluctuation, as we have already demonstrated.

Figure \ref{isospin_off-diagonal} represents the time evolution
of the two-point functions in isospin space in case i(a).
In this figure, the spatial average of their absolute value,
\begin{equation}
\langle | G_{a b} (t) | \rangle = \frac{1}{V}
\int |G_{a b}({\vec x},{\vec x},t)| d {\vec x},
\end{equation}
is shown. No event average is taken in Fig.  \ref{isospin_off-diagonal},
either. At the initial time, the two-point function is diagonal
in isospin space as in momentum space.
The off-diagonal components of the two-point function emerge
and develop as time elapses.
The absolute value of the off-diagonal
components evolves up to about one fourth of that of
the diagonal components by $t \sim 25$ fm, and then their
relative strength starts to decrease. 

\begin{figure}[t]
         \resizebox{80mm}{!}{\includegraphics{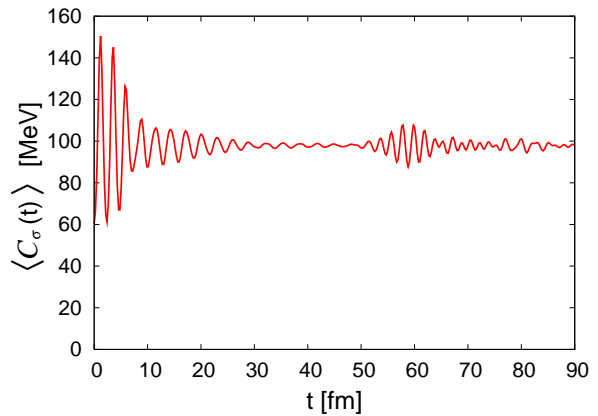}}
         \caption{(Color online) Time evolution of the spatial average of the
         sigma field.}
         \label{sigma-1}
\end{figure}

\subsection{Dissipation of the sigma field}

Finally, we explore whether and how the dissipation of the sigma field
is included in the approximation.
For this purpose, we chose the following initial
condition. The mean field of the sigma field $C_0$ was randomly
distributed according to the same Gaussian form as
Eq. (\ref{initial:mf4}) except that the first
equation was replaced by
$\langle C_{0} \rangle = 60$ MeV.
The pion fields and their conjugate
fields were set to zero at each lattice point. 
As for the quantum fluctuation
and correlation, the same initial condition
as in Section \ref{subsec_initcond},
i.e., those of independent coherent states 
with the degenerate mass $m_0 = 200$ MeV, were used.
Ensemble average was taken over 10 initial configurations
as before. In the calculation, both mode-mode and isospin-isospin
correlations are taken into account, i.e., case i(a).

\begin{figure}[t]
         \resizebox{80mm}{!}{\includegraphics{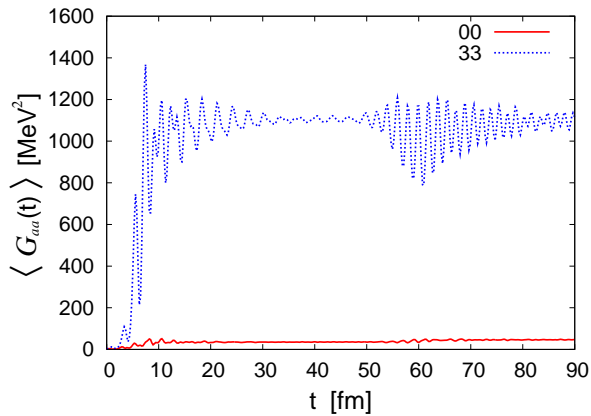}}
         \caption{(Color online) Time evolution of the quantum fluctuation of the
         sigma field (00) and the third component of the pion field (33).}
         \label{sigma-1f}
\end{figure}

Figure \ref{sigma-1} represents the time evolution of
the spatial average of the mean field part of the sigma field.
The result tells us that
the oscillation of the sigma field is damped quickly
and it approaches a constant.
Since initially the pion fields and their conjugate variables are all zero,
the mean field part of the pion fields remains zero throughout
the time evolution. On the contrary,
the quantum fluctuation of the pion fields is not zero initially and,
moreover, it is amplified substantially as shown for
the third component in Fig. \ref{sigma-1f},
while that of the sigma field changes only little. 
This amplification does not exist in the translationally invariant case.
Physically, the damping of the oscillation of the sigma field and the
growth of the quantum fluctuation of the pion fields correspond
to the decays of the sigma into two pions. As observed from Figs.
\ref{sigma-1} and \ref{sigma-1f}, the growth of the fluctuation
of the pion fields stops approximately at the same time
as the oscillation of the sigma fields ends. 

\section{\label{sec:conclusion}Conclusion}
In this paper, we have studied the dynamics of chiral phase
transition in spatially inhomogeneous systems with mode-mode
correlations in the framework of TDVA
with squeezed states for the 1+1 dimensional geometry.
Quantum effects and spatial inhomogeneity are both incorporated.

In the case with the mode-mode correlation, the
DCC domain formation continues beyond the
time scale of the rolling-down of the order parameter.
The quantum fluctuation is also amplified with approximately
the same time scale. This implies that squeezing of the states
takes place. On the other hand, no significant DCC domain formation or
squeezing of the states was observed 
in the case without the mode-mode correlation.
Thus, it is concluded that the mode-mode correlation
plays the key role in the formation of DCC domain.
This mode-mode correlation has not been taken into account
in preceding quantum calculations for DCC formation.
In addition, it was pointed
out that this mechanism is different from the previously
proposed parametric resonance amplification.

We have also examined the effect of the isospin-isospin
correlation on the domain formation of DCC for the first time.
We found that the isospin-isospin correlation makes
the DCC domain structure larger, but
its effect is less conspicuous than that of
the mode-mode correlation.
The isospin-isospin correlation has almost
negligible effects on the time evolution of the quantum fluctuation.
Thus, the isospin-isospin correlation does not lead to further
squeezing of the states.

We plan to include realistic geometry and dynamics such as
expansion, 
and carry out calculations in 2+1 or 3+1 dimension
in future work.
They are indispensable for the quantum mechanical
understanding of the DCC
formation in ultra-relativistic heavy ion collisions.

\begin{acknowledgments}
We thank T. Otofuji for making it possible
to use the computing facility at Tohoku Gakuin University.
N.~I. ackowledges the member of Nuclear Theory group at Yukawa Institute
for Theoretical Physics at Kyoto University for useful discussions and
encouragement.
M.~A. is partially supported by the Grants-in-Aid of the
Japanese Ministry of Education, Science and Culture, Grant Nos. 14540255
and 17540255, and Y.~T. is partially supported by the Grants-in-Aid of the
Japanese Ministry of Education, Science and Culture, Grant No. 15740156.
Numerical calculation was performed at
Yukawa Institute for Theoretical Physics at Kyoto University,
Tohoku Gakuin University, and Japan Atomic Energy Research Institute.
\end{acknowledgments}

\bibliography{apssamp}

\end{document}